# TOWARDS AUTOMATED VERIFICATION
# OF WEB SERVICES

Cátia Vaz
*INESC-ID Lisboa, ISEL-IPL*
*Rua Alves Redol 9, 1000-029 Lisboa*
*cvaz@cc.isel.ipl.pt*

Carla Ferreira
*INESC-ID, IST-UTL*
*Rua Alves Redol 9, 1000-029 Lisboa*
*carla.ferreira@dei.ist.utl.pt*

**ABSTRACT**

This paper proposes the use of model-checking software technology for the verification of workflows and business processes behaviour based on web services, namely the use of the SPIN model checker. Since the specification of a business process behaviour based on web services can be decomposed into patterns, it is proposed a translation of a well known collection of workflow patterns into PROMELA, the input specification language of SPIN. The use of this translation is illustrated with one business process example, which demonstrates how its translation to a PROMELA model can be useful in the web service specification and verification.

## 1. INTRODUCTION

In general we can think of business processes that make use of Web services and business processes that provide their functionality as Web services. Given the available technologies, e.g. WSDL and SOAP, Web services are described as self-contained modular business process applications. And thus, users are able to connect different components as complex workflows, i.e., a complex business process made of interconnected Web services. And, in order to define the business semantics of Web services, we must be able to specify and verify the execution of such workflows. Currently there are many efforts targeting the specification of the execution of Web services workflows, being the BPEL one of the best known efforts. However, we do not have tools which enable the verification of such workflows specifications.

In order to specify complex business processes it is important to identify generic and recurring constructs of workflow systems. A pattern based approach allows to express the workflow systems core characteristics in a way that is sufficiently generic for its application to a wide variety of technology offerings. So, using workflow patterns offers a language-independent and technology-independent means of expressing the behaviour of a business process. Russell et al (2006) presents a specification of the workflow patterns from the control flow perspective. Namely, activities and their execution ordering are described through different constructs, which allow flow of execution control such as sequence, choice, parallelism and synchronization.

Since workflow patterns are widely used in modeling the behavioural part of business processes, which is a domain that requires a certain degree of confidence, it is necessary to obtain a formal semantics for each workflow pattern. The application of formal methods to workflow patterns will allow formal verification of business processes. The formal verification can assure if the Web service workflow behaviour has or has not certain properties. Most of commercial workflow products does not have support for verification of Web services execution specification which may lead to models with errors and to undesirable executions of some or all instances of a given specification model. Some formal methods provide semantic analysis tools such as the SPIN model checker (Holzmann, 2003). With this tool, systems to be verified are described in PROMELA and the properties to be verified are expressed as Linear Temporal Logic (LTL) formulas.

This paper proposes the use of model-checking software technology for the verification of Web services workflows, namely the SPIN model checker (Holzmann, 2003). Since the specification of a workflow behaviour can be decomposed into patterns, it is proposed a translation of a well known collection of workflow patterns into PROMELA, the input specification language of SPIN. Namely, we have translated each pattern of the original set of twenty patterns, considering their reviewed definition (Russell et al, 2006). Some of the translated patterns will be presented within this paper. The use of this translation is illustrated with one Web service based business process scenario, which demonstrates how their translation to a PROMELA model can be useful in their specification and verification.

There are other works that consider workflow specification and verification using SPIN. Janssen et al (1998) has proposed a translation of a specification language of business process, Architectural Modelling Box for Enterprise Redesign (AMBER), into PROMELA. Kovács and Gönczy (2006) have proposed a method to check correctness properties of workflows implemented in BPEL. Dataflow networks are used to define the formal semantics of the workflow. The BPEL model is mapped into dataflow network and the dataflow network is mapped into a PROMELA model. Other work (Nakajima, 2002) has used SPIN for verifying Web Services Flow Language (WSFL) (Leymann, 2001) description. In particular, an encoding method that translates WSFL primitives to PROMELA is presented. These approaches are both focused on the translation of a workflow language to PROMELA, while this work is focused on the translation of workflow patterns to PROMELA, which are language and technology independent, in order to formally verify workflow systems.

An approach of giving a representation of the workflow patterns has been developed with π-calculus (Puhlmann and Weske, 2005), although it is not oriented towards automated verification. A benefit of using SPIN is in terms of visualizing counterexamples for negative results. Also, as the verification in π-calculus is done by checking bisimulation equivalence, some times results are not obtained in reasonable amount of time, even for the proofs of very simple correctness requirements (Song, H. and Compton, K., 2003). Other work has used a subset of π-calculus to model workflow patterns, Calculus of Communicating Systems (CCS) (Milner, 1995), but does not conform to standard CCS, and does not have a verification tool (Stefansen, 2005).

Wong and Gibbons (2007) has given a representation of the workflow patterns in Communicating Sequential Processes (CSP) (Roscoe, 1997). Comparing CSP and PROMELA, the latter is richer and strictly more expressive (e.g. asynchronous communication is supported and channels are first class objects in PROMELA but not in CSP). Furthermore, PROMELA's C-like syntax makes it more accessible to non-experts (Currie, 2000).

Yet Another Workflow Language (YAWL) (Aalst and Hofstede, 2005) was also used to represent workflow patterns. However, we believe that since SPIN is a model checker and PROMELA has a similar syntax to C it has an advantage over YAWL.

This paper is structured as follows. A brief introduction to the SPIN model checker is given in Section 2. Section 3 contains some of the definitions of workflow patterns and their translation to PROMELA. The other definitions of workflow patterns and their corresponding translations are presented in an extended version of this paper (Vaz and Ferreira, 2007). Two case studies are given in section 4 as well as the verification of some properties. Finally, in section 5 conclusions of this work are provided. This paper proposes the use of model-checking software technology for the verification of workflows and business processes behaviour based on web services, namely the SPIN model checker. Since the specification of a business process behaviour based on a web service can be decomposed into patterns, it is proposed a translation of a well known collection of workflow patterns into PROMELA, the input specification language of SPIN. The use of this translation is illustrated with two web services based business processes, which demonstrates how its translation to PROMELA models can be useful in web service specification and verification.

## 2. SPIN

The SPIN model checker is a tool to verify software systems developed by G. J. Holzmann (2003). SPIN provides a specification language, PROMELA, that describes the target system to be a collection of PROMELA processes with channel communications. The language allows dynamic creation of processes and both synchronous (rendezvous) and asynchronous communication through communication channels.

A SPIN model is a PROMELA model. Although called a program, it is more an executable model. A PROMELA program consists of variables, channels and processes. Processes are global objects, while variables and channels may be declared either as global or local to a process.

To create a process, one has to define a process template with the keyword `proctype`, and then use the `run` statement to create the process from the template. PROMELA also has the keyword `inline`, which behaves similar to macros and the inline functions in C++ although it has more restrictions. Often, several consecutive statements can be seen as one logical state change. It is possible to group these statements into an `atomic` or `d_step` block, which abstracts the statements into one state change.

Further, SPIN allows to express various properties in terms of linear temporal logic (LTL) (Manna and Pnueli, 1991) and to check if the program satisfies these LTL properties. When an error is found, SPIN reports it and shows the path of execution that led to the error.

In addition to model-checking, SPIN can also operate as a simulator, following one possible execution path through the system and presenting the resulting execution trace to the user.

## 3. WORKFLOW PATTERNS TRANSLATION

In this section it is introduced a translation into PROMELA of some of the workflow patterns proposed by Russell et al (2006). We focus on the translation of the generic workflow constructs. In this translation, processes, sub-processes and activities are mapped into PROMELA processes and control flow paths into PROMELA channels. The constructs are translated into `inline` definitions, which will be used in the description of control flow dependencies between activities and sub-processes in a workflow process. Messages between processes will be represented, without loss of generality, by integers in PROMELA.

In the translation of these patterns into the PROMELA language, we will use the following notation: `q` will represent a channel and `msg` the message sent or received in channel `q`; `qs` will denote an array of channels of `sizeq` and `msgs` an array of messages of `sizeq` to be sent or received in each channel in the array `qs`. The variable `choice` will be used to denote, when needed, the index of the channel in the array `qs` in which will be sent the message.

In some patterns, it is also necessary to consider the construct `myRun` that receives a process identification number `id` as a parameter and executes a new instance of the given process. The `id` of each process must be defined within the model. The construct `myRun` has also an instance identification `i`, which is used for activity instances in the multiple instance patterns. When there is no need to identify the new activity instance executed by the process `myRun`, the instance identification parameter will be -1.

The two workflow constructs that are widely used in the workflow patterns are the `send` and the `receive`. These constructs will be translated as `inline send(q, msg){ q!msg;}` and `inline recv(q, msg){ q?msg;}`, respectively. Note that variables are only declared in the scope of the proctypes that use these `inline` definitions.

With respect to the Basic Control Flow Patterns, we will present the translation of the patterns **Sequence**, **Parallel Split**, **Syncronization**, **Exclusive Choice** and **Simple Merge**.

**Sequence -** An activity `B` in a workflow process is enabled after the completion of a preceding activity `A` in the same process. The process `A` in PROMELA should use the `send` definition and the process `B` should use the `recv` definition.

```
chan q = [1] of {int};
proctype A(){
  /* Do work. */
  send(q,1); /* To activate process B. */ }

proctype B(){
  int x;
  recv(q,x); /* Waiting token. */
```

/* Do work. */}

Note that the comments /* Do work. */ in the above processes should be replaced by the details of each activity.

**Parallel Split -** This pattern is defined as being a mechanism that will allow activities to be performed concurrently, rather than serially. A single path through the process is split into two or more paths so that two or more activities will start at the same time. This pattern is translated by the following `inline` definition, where each channel in the array `qs` is used to communicate with each activity.

```
inline parallelSplit(qs, sizeq, msg){
  int n;
  n=0;
  atomic {
    do
      :: n<sizeq -> qs[n]!msg[n]; n++;
      :: n>=sizeq -> break;
    od; }}
```

The process in PROMELA which represents the activity that splits the process must use the `parallelSplit` definition and the processes which represent the activities to be initiated must use the `recv` definition.

**Synchronization -** The Synchronization pattern combines the paths that were generated by the **Parallel Split** pattern. The final set of activities within the flows must be completed before the process can continue. This pattern is translated to the following `inline` definition.

```
inline synchronization(qs, sizeq, msgs){
  int n, count;
  n=0; count=0;
  /* MAXARRAYSIZE: The capacity of the arrays defined in the file
   * which contains the translations of the workflow patterns */
  int aux[MAXARRAYSIZE];
  do
    ::n<sizeq -> aux[n]=0; n++;
    ::n==sizeq -> n=0; break;
  od;
  skip;
  S:
    if
      ::((aux[n]==0) && (len(q[n]) > 0) && count<sizeq)->
          aux[n]=1; q[n]?msg[n]; count++
      ::count>=sizeq -> goto E
      ::else -> skip;
    fi;
  n++;
  if
    ::n==sizeq -> n=0; timeout;
    ::n<sizeq -> skip;
  fi;
  goto S;
  E: skip;}
```

The process in PROMELA that receives the input branches must use the above definition and the processes to be synchronized must use the `send` definition. It is denoted by `aux`, an auxiliary array of size `sizeq` to distinguish between the activities already completed from the others. After finishing, each activity sends a message through a channel in the array `qs` to report it (e.g. if the activity that communicates through channel `qs[n]` is finished, it will send a message through this channel reporting that). In this pattern translation, all the channels in the array `qs` are transverse to see if there is something to receive from each one of them. If there is, it will be received and marked in the array `aux` (e.g. `aux[n]=1`). Thus, the process in PROMELA which use the `synchronization` definition will only continue if it has received a message from each channel. The use of the keyword `timeout` is to avoid process starvation, giving the opportunity to other processes to execute.

**ExclusiveChoice -** This pattern is defined as being a split of the control flow into two or more exclusive alternative paths. The pattern is exclusive in the sense that only one of the alternative paths may be chosen for the process to continue. This pattern is translated by the following `inline` definition.

```
inline exclusiveChoice(qs, sizeq, choice, msg){
  if :: (choice>=0 && choice<sizeq) -> qs[choice]!msg;
     :: else -> skip;
  fi;}
```

The process representing the activity which makes the choice must use this definition, and the alternative processes must use the `recv` definition.

With respect to the Cancellation Patterns, we will present the translation of the **Cancel Activity** and **Cancel Case** patterns. To translate these it is necessary that each PROMELA process that may be canceled have a specific channel for this purpose. Thus, if the process receives a message from that channel, it should terminate. We will denote a single cancel channel as `qCancel` and an array of canceling channels as `qsCancel`.

**Cancel Activity -** The Cancel Activity pattern provides a mean of withdraw an enabled activity before starting to execute. However, if the activity has already started, it is disabled and, where possible, the currently running instance is halted. To translate this pattern it is necessary that the PROMELA process which represents the activity to be canceled includes an escape sequence as follows.

```
unless { len(qCancel)>0; skip; }
```

**Cancel Case -** This pattern describes the situation where is necessary to remove a complete process instance. This includes executing activities, those which may execute at some time and all sub-processes. More generally, this may be used to cancel individual activities, regions or the whole workflow instance. The PROMELA model needs to have a global array `piIds` representing the relations between the processes, e.g., the parent of the process with identification number `x` is the process with identification number `piIds[x]`. In the following `inline` definition, `id` is the identification number of the process where this inline definition is being included.

```
inline cancelCase(qsCancel,sizeq,piIds,msgs,id){
  int i=0;
  do :: i<sizeq && piIds[i]==id -> qsCancel[i]!msgs[i];i++;
     :: i==sizeq ->break;
     :: else -> i++;
  od;}
```

To implement this pattern, the process and all its activities and sub-processes must include an escape sequence as follows.

```
unless{ len(qsCancel[id])>0; cancelCase(qsCancel,sizeq,piIds,msgs,id)}
```

Each process that receives a message from its canceling channel `qsCancel[id]`, sends a canceling message, before terminate, to each one of its sub-processes or activities in order to cancel them.

## 4. CASE STUDY

In order to ensure the reliability of business process, formal verification methods are needed. This section shows how the above translation of workflow patterns can be useful for the formal verification of business process models. A standard example of business process is the Travel Agency. In this section it will be illustrated how to apply the translation described above to this business process examples and, subsequently, a property of these process will be checked. It will be also illustrated how this business process can be modeled in Business Process Modeling Notation (BPMN) (OMG, 2006).

### 4.1 Travel Agency

This case study consists of a simple travel agency where customers can book a trip. The process of booking trips involves booking a flight and a hotel. If both bookings succeeds, the payment follows. Otherwise, the booking of the trip is canceled. A description of this process can also be found in (OMG, 2006).

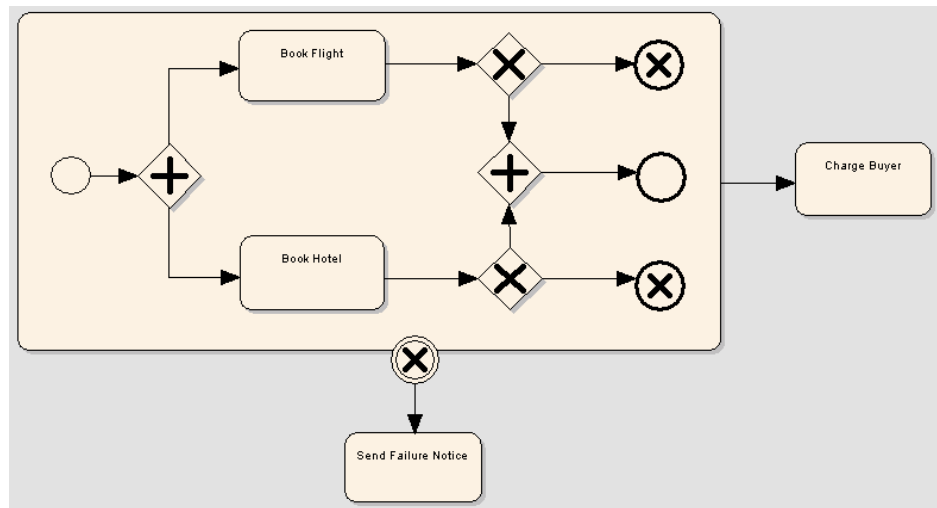

Figure 1. A BPMN diagram for the Travel Agency example

Figure 1 shows the business process modeled as a BPMN diagram. The process of booking is represented by the transaction **Book** and executes in parallel two activities: the activity **Book Flight** and **Book Hotel**. If both activities succeed, the activity **Charge Buyer** follows, otherwise, the process **Book** is canceled and follows the activity **Send Failure Notice**.

### 4.2 The PROMELA Model

In what follows, the activities involved in Travel Agency example will be specified as PROMELA processes. The PROMELA model will have two global channels. The array of channels denoted by `qs` is used for communication between the process **Book** and its sub-processes **Book Flight** and **Book Hotel**. However, in the situation of canceling, the communication among these processes is made by their respective canceling channels, denoted by `qsCancel`. The other two channels, `q1` and `q2`, are used by process **Book** to communicate, when needed, with processes **Failure Notice** and **Charge Buyer**.

We denote by `qs1` and `qs2` the arrays of channels defined within the processes **Book Flight** and **Book Hotel**, respectively. Each one of these auxiliary array of channels include one channel of `qs` and one channel of `qsCancel`.

**Book -** This PROMELA process translates the process of booking.

```
proctype Book(){
{ int ids=1; /* The id of this process is 1. */
  piIds[ids]=0;
  int msgs[4];
  /* Receive personal information of the costumer. */
  ids++; piIds[ids]=1;
  run myRun(2, -1); /* Run BookFlight. */
  ids++; piIds[ids]=1;
  run myRun(3, -1); /* Run BookHotel. */
  parallelSplit(qs,2,1);
  synchronization(qs,2,msgs);
  send(q1,1);
} unless { len(qsCancel[1])>0;
          cancelCase(qsCancel,4,piIds,msgs,1); send(q2,1); }}
```

**Book Flight -** This activity represents the booking of a flight. In this activity is decided if it is possible to book a flight according to the interests of the costumer. If it is not possible, the activity **Book Flight** is canceled and this information is sent to the process **Book** in order to the whole process be canceled. Since the choice of one of the possibilities depends on the details of the business process, we choose non-determinalistically, without loss of generality to the verification, one of the three possibilities.

```
proctype BookFlight(){ /* 2 is the id of this process. */
{ int x, msgs[4];
  chan qs1[2];
  qs1[0]=qs[0]; qs1[1]=qsCancel[1];
  recv(qs[0],x); /* Waiting token. */
  /* Decide if it is possible or not to book a flight. */
  if
    ::x=0 /* Not to cancel. */
    ::x=1 /* To cancel. */
  fi;
  exclusiveChoice(qs1,2,x,1);
} unless {len(qsCancel[2])>0; cancelCase(qsCancel,4,piIds,msgs,2); } }
```

**Book Hotel -** This activity represents the booking of a hotel. In this activity is decided if it is possible to book a hotel suitable for the interests of the costumer. If it is not possible, the activity **Book Hotel** is canceled and this information is sent to the process **Book** in order to the whole process be canceled. Once again, we choose non-determinalistically, without loss of generality to verification, one of the two possibilities.

```
proctype BookHotel(){ /* 3 is the id of this process. */
{ int x, msgs[4];
  chan qs2[2];
  qs2[0]=qs[1]; qs2[1]=qsCancel[1];
  recv(qs[1],x); /* Waiting token. */
  /* Decide if it is possible or not to book a hotel. */
  if
    ::x=0 /* Not to cancel. */
    ::x=1 /* To Cancel. */
  fi;
  exclusiveChoice(qs2,2,x,1);
} unless { len(qsCancel[3])>0; cancelCase(qsCancel,4,piIds,msgs,3); } }
```

**Charge Buyer -** This activity is activated by the process **Book** in order to charge a payment to the costumer that has requested the booking.

```
proctype ChargeBuyer(){
  int x; recv(q1,x);
  /* Charge the costumer. */}
```

**Send Failure Notice -** This activity is activated by the process **Book** and must inform the costumer that the booking process has failed.

```
proctype sendFailure(){
  int x; recv(q2,x);
  /* Send failure notice to the costumer. */}
```

Bellow is the corresponding PROMELA model of the BPMN diagram in figure 1.

```
/* File with the translations of the workflow patterns. */
#include "utils.pr"

chan qs[2] = [1] of {int};
chan q1= [1] of {int}; chan q2= [1] of {int};
chan qsCancel[4]= [1] of {int};
int piIds[4];

proctype myRun(int id, int n){
  if
    ::(id==1) -> run Book()
    ::(id==2) -> run BookFlight()
    ::(id==3) -> run BookHotel()
  fi;}

proctype Book(){...}
proctype ChargeBuyer(){...}
proctype SendFailure(){...}
proctype BookFlight(){...}
proctype BookHotel(){...}
init{ atomic{ run Book(); run ChargeBuyer(); piIds[0]=-1; run SendFailure();}
}
```

## 4.2 Properties Verification

It is possible to check if the Travel Agency business process model satisfies certain properties using the SPIN model checker. These properties are expressed using Linear Temporal Logic (LTL). A LTL formula, which is used to specify properties, is built up from a set of propositions variables, combined using boolean connectives and/or temporal operators. In the following, formulas will be represented by `f, f1, g1, h1,...`, and propositions by `p, q, r, p1, q1, r1,...`. Boolean conjunction, disjunction, negation and implication of formulas are denoted by `f1 && f2, f1 || f2, ! f` and `f1 → f2`, respectively. The temporal operators eventually, always, until and next are denoted by `<> f, [] f, f1 U f2` and `X f`, respectively. For instance, the formula `[] f` means that `f` will be satisfied in all states in the future and the formula `f1 U f2` states that `f2` will certainly hold in the future and `f1` will continuously hold until then.

Some properties which the Travel Agency business process model should satisfy can be expressed in a LTL formula and checked with SPIN. We will show two properties, which corresponds to typical temporal properties, objective (`[] (p1 → <> ( p2 || p3))`) and response (`[] ( p1 → ( <>p2 ))`). We will express these properties using a global `int` variable `s`, in order to express the state of the of the PROMELA model execution. In the beginning of the execution, `s` will have the value 0. The other possible values are 1, 2, 3, 4 and 5 to represent the end states of the processes `ChargeBuyer, SendFailure, Book, BookFlight` and `BookHotel`, respectively.

The objective property is that whenever we invocate this business process, the process either charge the buyer or send him a failure notice to tell him that the booking was not possible. By making the following definitions `#define p (s==0), #define q (s==1)` and `#define r (s==2)`, it is possible to formally express this property in LTL as `[] (p -> <> (q || r))`.

Finally, a response property which the model should satisfy is that if both hotel and flight are booked then the buyer will be charged. By making the following definitions `#define p (s==4)`, `#define q (s==5)` and `#define r (s==1)`, it is possible to formally express this property in LTL as `[] (( p && q) → <>(r))`.

Both properties were verified with the SPIN model checker, which confirms the intended behaviour of the model specification.

## 5. CONCLUSION

This paper proposes the use of model-checking software technology for the verification of workflows and business processes behaviour based on web services, namely the SPIN model checker. Since the specification of a business process behaviour based on a web service can be decomposed into patterns, it is proposed a translation of a well known collection of workflow patterns into PROMELA. This translation is applied to a case study, namely the Travel Agency example. It is also illustrated the verification of two properties. Thus, this simple example demonstrate how PROMELA models can be useful in web service specification and verification.

Future work will concern an automatic translator of BPMN models (or models described in other process modeling languages) to PROMELA models. It is also interesting to further express required properties of the workflow patterns in linear temporal logic in order to verify them with SPIN model checker.